# Current Leads, Links and Buses

*A. Ballarino[1]*
CERN, Geneva, Switzerland

**Abstract**
Electrical transfer from a room temperature power source to a superconducting system is done via conventional or superconducting current leads and superconducting buses or links. The principles of optimization of these devices are presented, with emphasis on the cryogenic, electrical, and superconductor related aspects that drive choices for a system.

## 1 Introduction

The powering of a superconducting system requires transfer of current from ambient temperature, where the current is generated, to the cryogenic environment. This electrical transmission, which takes place inside a cryostat, is performed by the current leads, i.e. devices that transport current in a significant gradient of temperature and represent, therefore, one of the major sources of heat inleak into the cryostat.

Designing cryogenic devices requires precise knowledge of material properties at the operating temperatures, some of which may vary by orders of magnitude between ambient and cryogenic conditions. Thermal conductivity in metals and alloys spans across several orders of magnitude and is dominated by thermal conduction by electrons, scattered by lattice phonons and imperfections. Electrical resistivity also varies by several orders of magnitude depending upon the type of material, degree of alloying, and purity. It decreases with temperature and it shows a typical plateau corresponding to a residual resistivity value due to electron scattering by impurities and lattice imperfections. Metals follow the Wiedemann–Franz–Lorenz (WFL) law, which establishes an inverse proportionality between thermal conductivity ($k$) and electrical resistivity ($\rho$) at a given absolute temperature ($T$):

$$k(T)\rho(T) = L_0 T \tag{1}$$

where $L_0$ is a constant, the Lorenz number, which was calculated by Sommerfeld in 1927 using Fermi–Dirac statistics:

$$L_0 = \frac{\pi^2}{3}\left(\frac{k_B}{e}\right)^2 = 2.45 \cdot 10^{-8} \text{ W}\,\Omega\,\text{K}^{-2} \tag{2}$$

where $k_B$ is Boltzmann's constant and $e$ the electron charge [1].

Good electrical conductors are also good thermal conductors, and the transfer of electrical current in metals that operate in a gradient of temperature, for which one needs to minimize both ohmic losses and thermal conduction, is bound by the existence of a minimum heat inleak into the cryogenic environment. For current leads operating between ambient and liquid helium temperature, this minimum heat inleak ($Q_{C,\min}$) per unit current is about 47 W kA$^{-1}$ for a conduction-cooled lead and 1.1 W kA$^{-1}$ for a gas-cooled lead operated in self-sustained mode (self-cooled lead). This value is independent of material properties, and it represents an optimum performance in terms of minimum heat inleak associated with the transmission of a given current inside a cryostat. For a cryogenic system, minimum heat inleak means minimum liquid cryogen consumption or minimum refrigeration power load. The achievement of this optimum performance is the main objective of the design of a current lead.

---

[1] amalia.ballarino@cern.ch

The WFL law is a useful correlation that can be used to derive the thermal conductivity of metals as a function of temperature from the electrical resistivity, the latter being easier to measure with precision. However, it is known that highly-alloyed materials depart from the WFL law, and in many metals at low temperature the Lorenz number deviates from the theoretical Sommerfeld value [2]. This means that for the optimization of a current lead one should preferably use the measured temperature-dependent material properties instead of the WFL relationship. The value of the minimum heat inleak only changes by a few per cent, but the use of the material properties enables correct definition of the specific geometry that enables the achievement of the optimum performance.

Before the discovery of High-Temperature Superconductors (HTS), current leads were made from metals or alloys. Design optimization in this case is mainly associated with the calculation of the optimum geometry of the lead for a selected material and for a given current to be transferred. Discovery of high-temperature superconductivity and production of technical HTS opened an innovative way to improve the performance of conventional current leads by incorporating in them materials that escape the WFL law in view of their low thermal conductivity and ideal zero resistivity in the superconducting state. HTS current leads, also called binary leads, enable the reduction of the heat inleak into the liquid helium bath by a factor of up to ten with respect to the minimum obtainable with conventional leads [3]. The corresponding total saving in cooling power is about 30% [3].

Studies on the possibility of integrating HTS materials in leads started in the 1990s [4–6], when prototypes were also made [7–10], and the first large-scale application of HTS current leads was with the Large Hadron Collider (LHC) machine, where more than 1000 HTS leads are used to feed the superconducting magnet circuits of the accelerator [11]. Thanks also to the developments carried out for this large-scale application, HTS leads have today become a recognized standard technology. The International Thermonuclear Experimental Reactor (ITER) will use HTS current leads for feeding the superconducting coils of the tokamak being built at Cadarache, France [12].

## 2    Conventional current leads

Depending on the cooling mode, conventional current leads can be classified into conduction-cooled or gas-cooled (Fig. 1). In the first case, the lead operates in vacuum, while in the second case there is convective heat transfer. If the heat conducted by the lead into the liquid cryogen bath generates the gas that is then used for cooling the lead itself, the gas-cooled lead is termed self-cooled. In all cases, the optimization requires the solution of the steady-state heat balance equation that includes energy generation due to the ohmic heating associated with the transfer of current.

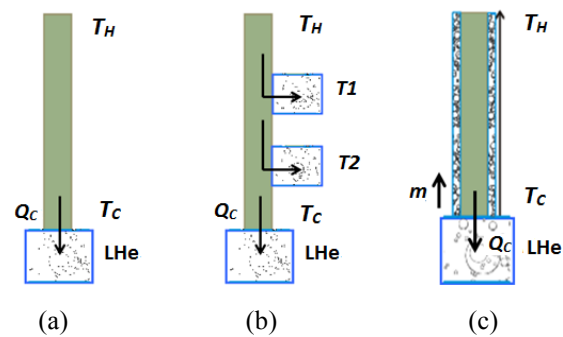

**Fig. 1:** Cooling scheme for a conventional current lead: (a) conduction-cooled current lead without heat sinks; (b) conduction-cooled current lead with intermediate heat sinks; (c) self-cooled current lead.

## 2.1 Conduction-cooled conventional current leads

Under the assumption that the total length of the lead ($L$) is much larger than its cross-section ($A$), the mono-dimensional heat balance equation can be considered. The minimization of the heat inleak at the cold end of a conventional lead ($Q_C$) is obtained by imposing a zero slope in the temperature profile ($T(x)$) at ambient temperature ($x = L$), which corresponds to zero thermal conduction at the warm end of the lead.

The minimum heat inleak ($Q_{C,\min}$) per unit current ($I$) is almost independent of material properties. This can be easily demonstrated for a conduction-cooled lead under the assumption that it follows the WFL law. From the solution of the mono-dimensional steady-state heat balance equation:

$$\frac{d}{dT}\left(k(T)A\frac{dT}{dx}\right) + \rho(T)\frac{1}{A}I^2 = 0 \tag{3}$$

introducing:

$$Q(T) = k(T)A\frac{dT}{dx} \tag{4}$$

one can calculate $Q_C$ per unit current transferred:

$$\frac{Q_C}{I} = \sqrt{Q_H^2 + L_0 I^2(T_H^2 - T_C^2)}, \tag{5}$$

which is minimized when the heat conducted at the warm end of the lead ($Q_H$) is zero:

$$\frac{Q_{C,\min}}{I} = \sqrt{L_0 I^2(T_H^2 - T_C^2)}. \tag{6}$$

The geometry that enables the achievement of optimum performance is dependent upon material properties and it can be obtained from:

$$\left(\frac{L}{A}\right)_{\text{opt}} = \int_{T_C}^{T_H} \frac{k(T)}{\sqrt{(Q_{C,\min}^2 - L_0 I^2(T^2 - T_C^2))}}. \tag{7}$$

From Eq. (7) it appears that, for a given material, the optimum thermal performance at a given current corresponds to a constant length-to-cross-section ratio. The optimum geometry of a lead is often conveniently reported in terms of shape factor (SF), which applies to all optimized leads made from a specific material, regardless of current:

$$\text{SF} = \left(\frac{IL}{A}\right). \tag{8}$$

The mono-dimensional temperature profile $T(x)$ of an optimized conduction-cooled lead can be obtained from numerical integration of Eq. (7) for a given operating current and a selected geometry, i.e. length and corresponding cross-section. The total length of a current lead is often imposed by the requirements from integration inside a cryostat, and the cross-section is derived from Eq. (8).

It should be noted that while $Q_{C,\min}$ at the current for which the lead is optimized does not depend on material properties, the heat conducted by the lead when it is either in stand-by mode ($I = 0$ A) or it operates at a current lower than nominal does depend on material properties. Leads made from materials with high thermal conductivity, e.g. copper with Residual Resistivity Ratio (RRR) above 100, in stand-by conditions conduct, inside the cryogen bath, significantly higher heat than those made from materials with low thermal conductivity, e.g. phosphorus deoxidized copper (RRR < 10) or brass. This is an important input for the selection of the material, in particular for systems that are not continuously powered. Considerations of stability in transient conditions also favour the selection of materials with lower thermal conductivity, at the cost of a larger cross-section.

An optimized conduction-cooled current lead operating between room temperature and the liquid helium bath has a heat inleak at 4.2 K of about 47 W kA$^{-1}$. Such a high thermal load limits the application of this type of lead to systems operated at low currents. Improvement of thermal performance can be obtained by additional extraction of heat at higher temperatures using intermediate heat sinks (Fig. 1(b)), which are provided either by cold stages of cryo-coolers or by dedicated gas-cooled heat exchangers.

Conduction-cooled current leads with intermediate heat sinks were developed for the powering of the LHC orbit corrector magnets. These circuits operate at a maximum current of 60 A. A hybrid conductor consisting of a tapered copper-plated brass rod, electrically insulated and enclosed in a stainless steel tube, transfers the current across the technical vacuum from room temperature to the 1.9 K superfluid helium bath where the magnets are operated [13]. Heat sinking is provided by two heat exchangers cooled by helium gas available in the LHC cryogenic system at about 50 K and 20 K, respectively (Fig. 2). With this design, the heat inleak at 60 A is reduced to about 0.19 W, to be compared with the 2.8 W that would be conducted by a lead operating at the same current and with no intermediate heat sinking.

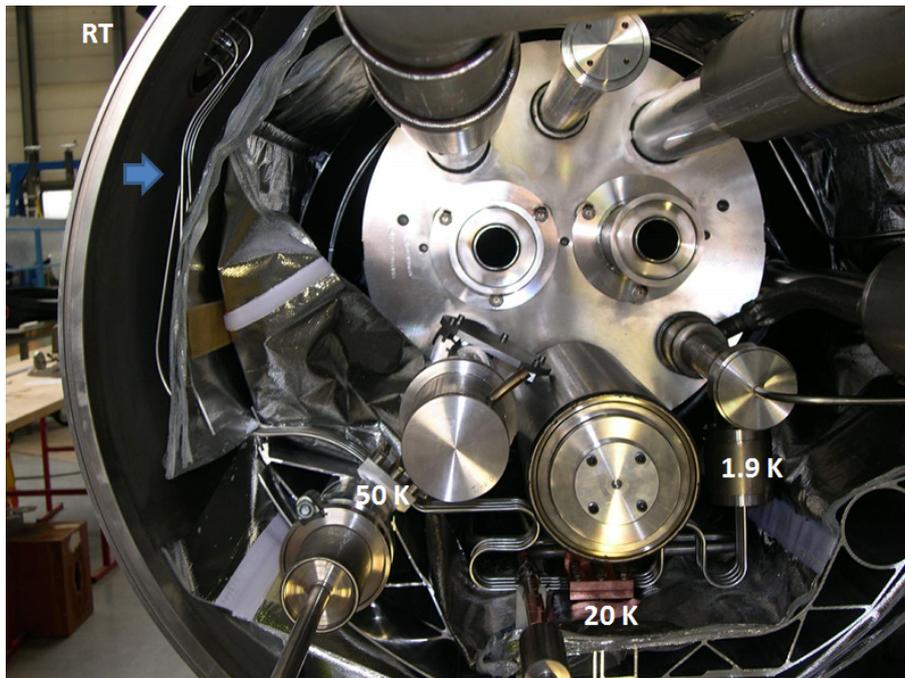

**Fig. 2:** LHC orbit corrector conduction-cooled current leads (path indicated by arrow) integrated inside the LHC cryostat. The leads operate in the cryostat vacuum insulation between room temperature (RT) and 1.9 K. They are thermalized against two heat sinks cooled by He gas, available in the LHC cryogenic system, which are maintained at about 50 K and 20 K.

Conduction-cooled current leads are often chosen for their simplicity, provided the cryogenic system can afford the associated thermal losses. They do not require an active control of flow and they can have, for low current ratings, a compact design that incorporates mechanical flexibility to enable easy integration in a cryostat [13]. Drawbacks are the higher heat inleak and, for systems that include intermediate heat sinks, the difficulty of providing a good thermalization, in the vacuum insulation of the cryostat, at the location where the heat is extracted from the electrically insulated conductor.

## 2.2 Self-cooled conventional current leads

The optimization of a self-cooled conventional current lead requires the solution of a system of differential equations expressing the heat balance between the conductor and the coolant. In a steady state and in the mono-dimensional case, neglecting thermal conduction in the gas, the system is:

$$\frac{d}{dx}\left(k(T)A\frac{dT}{dx}\right) = -\frac{\rho(T)I^2}{A} + Ph(T)(T - \vartheta),$$

$$mc_p(\vartheta)\frac{d\vartheta}{dx} = Ph(T)(T - \vartheta) \tag{9}$$

where $T$ is temperature of the conductor, $k$ and $\rho$ are the conductor thermal conductivity and electrical resistivity, $A$ and $P$ are the conductor cross-section and wetted perimeter, $\vartheta$ is the temperature of the gas, $c_p$ is the specific heat of the gas, $m$ is the gas mass flow rate, and $h$ is the heat exchange coefficient between conductor and coolant.

In self-cooled conditions, the heat conducted into the liquid bath ($x = 0$) generates the mass flow rate needed for cooling the lead:

$$m = \frac{kA}{C_L}\frac{dT}{dx} \tag{10}$$

where $C_L$ is the latent heat of vaporization of the liquid cryogen.

The solution of Eq. (9) requires the knowledge of the current lead geometry for the definition of $P$ and $h$. A first optimization of the lead can be made by assuming a perfect heat exchange between conductor and gas ($h \to \infty$). With this assumption, the heat balance equation becomes:

$$\frac{d}{dx}\left(k(T)A\frac{dT}{dx}\right) + \frac{\rho(T)I^2}{A} - mC_p(T)\frac{dT}{dx} = 0. \tag{11}$$

Taking as boundary conditions the temperatures at the warm end ($x = L$, $T = T_H$) and at the cold ($x = 0$, $T = T_C$) end of the lead and applying Eq. (10), it is possible to calculate from Eq. (11) the minimum heat inleak per unit current [14] and the corresponding temperature profile characterized by a zero slope at the warm end of the conductor ($x = L$). For a self-cooled lead operating between room temperature and liquid helium, the minimum heat inleak is about 1.1 W kA$^{-1}$. As for a conduction-cooled lead, this optimum performance is almost independent of material properties, and the geometry of the lead is defined by an optimum SF. Figure 3 reports the heat conducted at 4.2 K by two self-cooled leads, one made from stainless steel and the other from pure copper, as a function of the SF [15]. The minimum heat inleak is about 1.1 W kA$^{-1}$ in both cases, while the optimum SF differs because of different material properties. Moving away from the optimum SF implies always higher heat inleaks and higher mass flow rates for cooling. Shape factors smaller than the optimum correspond to leads that operate in an over-cooled regime – this is the case, for instance for leads that have a too-large cross-section for a given current and length – while an SF smaller than optimum corresponds to leads that are sub-cooled. This latter case is critical in that leads having a too-small cross-section for a given current and length suffer thermal runaway – steady-state operation with local temperatures higher than room temperature is only possible within a small range of geometries.

Since the SF depends on the properties of the material, use should be made of the real conductor properties as a function of temperature over the full range of operation. It is clear from Fig. 3 that leads made from pure metals are more sensitive to variations of geometry around the optimum. The final optimization of a lead has to be done via the solution of Eq. (9) taking into account the detailed geometry selected in the design phase.

Table 1 summarizes the minimum heat inleak per unit current of conventional conduction-cooled and gas-cooled current leads operated between room temperature and either liquid helium or liquid nitrogen temperature. In Fig. 3 is reported the temperature profile calculated and measured on self-cooled current leads optimized for transferring 18 kA between room temperature and the liquid helium bath [15].

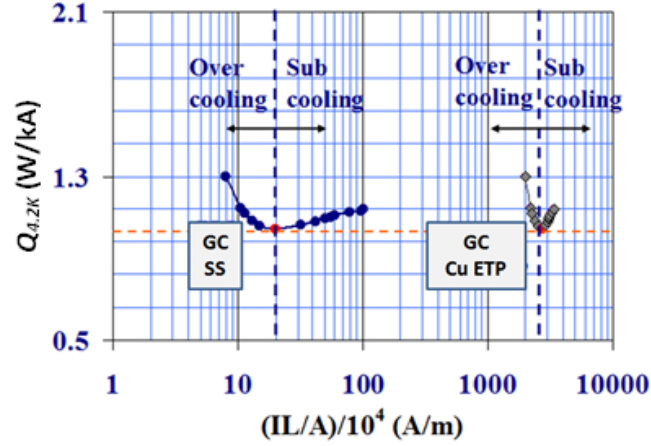

**Fig. 3:** Dependence of heat inleak on shape factor for two gas-cooled (GC) conventional current leads made from stainless steel (SS) and copper (Cu ETP) [15]. The leads transfer current to a liquid helium bath. The minimum heat inleak is the same while the optimum SF differs because of different material properties.

Table 1: Minimum heat inleak (W kA$^{-1}$) of conventional current leads

| Type of lead | $Q_{C,min}$ (4.2 K) | $Q_{C,min}$ (77 K) |
| --- | --- | --- |
| Conduction-cooled | 47 | 45 |
| Gas-cooled | 1.1[a] | 23[b] |

[a]Helium gas cooling; [b]nitrogen gas cooling

## 3   HTS current leads

A HTS current lead consists of a resistive part, made from a normal conductor, and a superconducting part made from a high-temperature superconducting material (Fig. 5). The resistive part operates between room temperature ($T_H$) and the maximum operating temperature of the HTS ($T_{HTS}$), while the HTS part provides the electrical transfer in the temperature region from $T_{HTS}$ to the cryogenic environment ($T_C$). The use of HTS materials offers the potential for reducing the refrigeration requirements to values significantly lower than those achievable with conventional optimized current leads. The total gain in terms of reduction in the exergetic cost of the refrigeration depends on the final design of the HTS current lead. The latter is influenced by the availability of cryogens for the choice of the cooling mode, and by the electrical requirements of the superconducting system, which include the definition of protection needs and the conditions to be met in transient operation.

The most efficient cooling schemes for HTS leads are based on independent cooling of the resistive and HTS parts. For a self-cooled HTS current lead relying entirely on cooling via a mass flow generated by the heat conducted by the lead itself into a helium bath, the reduction in exergetic cost of the refrigeration with respect to a conventional current lead is limited to about 15%. This is because the cooling is driven by the resistive part, which determines the helium boil-off requirements [4], while the HTS part is over-cooled. Self-cooling using a nitrogen bath or forced flow cooling by helium gas are both efficient options for cooling the resistive part of an HTS lead. The reduction in total cooling power obtained by adopting these cooling schemes is up to about 30% of a conventional self-cooled lead, while the reduction in heat conducted into the helium bath is a factor of up to 10 [11]. Conduction cooling is limited to applications that usually rely on cryo-coolers and run at low currents ($I \leq 1.5$ kA). In all cases, the HTS part of the lead can be either conduction-cooled or gas-cooled.

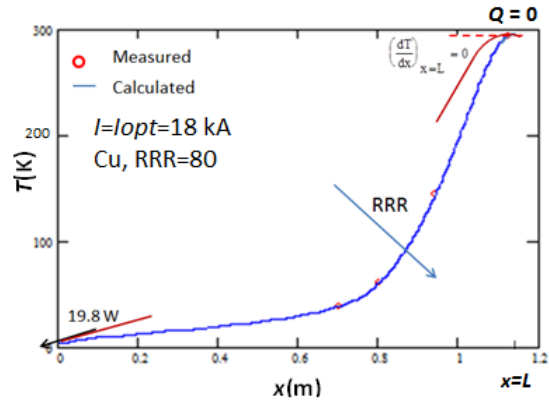

**Fig. 4:** Temperature measured and calculated on a CERN self-cooled current lead, 1.2 m long, optimized and operated at 18 kA [15]. The upper curve indicates the qualitative change in the optimum temperature profile corresponding to a change in material properties – Cu with lower RRR.

The optimization of a HTS current lead consists of two parts: the optimization of the resistive part and of the superconducting part (Fig. 5).

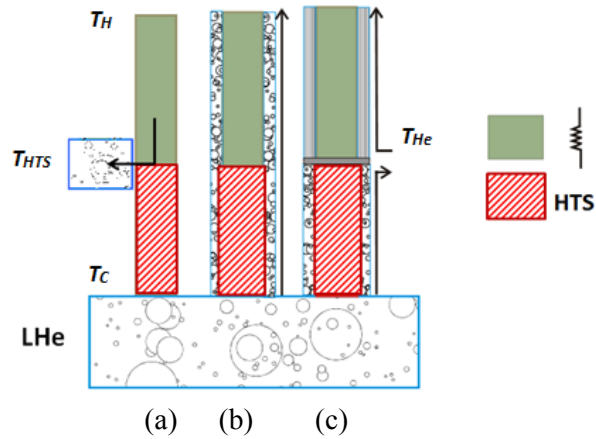

**Fig. 5:** Cooling scheme of: (a) HTS conduction-cooled current lead; (b) HTS self-cooled current lead; and (c) HTS gas-cooled current lead. $T_{HTS}$ is the temperature at the top end of the HTS part and $T_{He}$ is the inlet temperature of the gas cooling the resistive part of the lead.

### 3.1 HTS current leads: resistive part

The optimization of the resistive part of an HTS lead requires the solution of Eq. (9) according to the description in Section 2. If self-cooling is provided by nitrogen boil-off, optimization is identical to that of a conventional self-cooled lead and the performance is as reported in Table 1.

Cooling via forced flow of helium gas opens possibilities for different operating temperature ranges and optimization schemes [11]. As opposed to self-cooled conventional current leads, where the helium mass flow rate is bound to the heat load by Eq. (10) and where the bottom end of the lead is at the same temperature of the liquid cryogen, in the gas-cooled resistive part of a HTS lead $T_{HTS}$ can assume a wide range of operating values. The latter depend on the inlet temperature of the coolant ($T_{He}$) and on the maximum operating temperature of the superconductor. In the case of cooling with helium gas, $T_{HTS}$ can assume any value in the range up to about 85 K. It can be shown [15] that also in this case the minimum mass flow rate required for the cooling of the conductor is the one that gives a temperature profile characterized by a zero slope at room temperature. As for conventional self-cooled current leads, for a given $T_{He}$ there is an optimum SF. The SF depends both on material properties and on $T_{HTS}$.

Figure 5 maps the optimum performance of the helium gas-cooled resistive part of a HTS lead optimized for different operating conditions [11]. The minimum mass flow rate per unit current is calculated for a wide range of inlet temperatures of the coolant and of operating temperatures of the HTS. The performance and the operating conditions of the LHC and of the ITER HTS current leads are also reported. In the LHC, cooling is done via helium gas entering at about 20 K, and the top end of the HTS is maintained at 50 K [9]. In the case of the ITER HTS leads, cooling is done via helium gas entering at about 50 K, and the top end of the HTS is maintained at 65 K [12]. The inlet temperature of the gas depends on the cryogenic system. The operating temperature of the HTS is the result of an optimization that takes into account that, for a given $T_{He}$, higher $T_{HTS}$ requires lower mass flow rates but a larger cross-section of superconductor. There is a trade-off, which is a system choice.

Figure 7 reports the SF for each of the operating points in Fig. 6. The optimum geometry of the resistive part of the lead is calculated using the material properties of copper with RRR = 70. The SF depends on material properties and on operating conditions ($T_{HTS}$ and $T_{He}$).

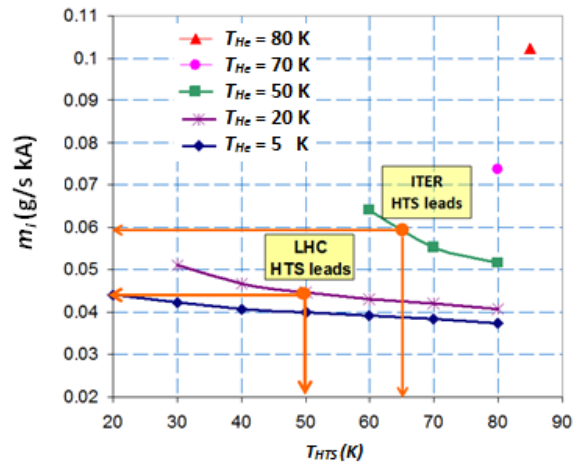

**Fig. 6:** Minimum mass flow rate per unit current ($m_I$) required for cooling the resistive part of a HTS lead as a function of the operating temperature of the HTS ($T_{HTS}$) and of the inlet temperature of the helium gas ($T_{He}$). Operating conditions and optimum performance of the resistive part of the LHC and ITER HTS current leads are also reported.

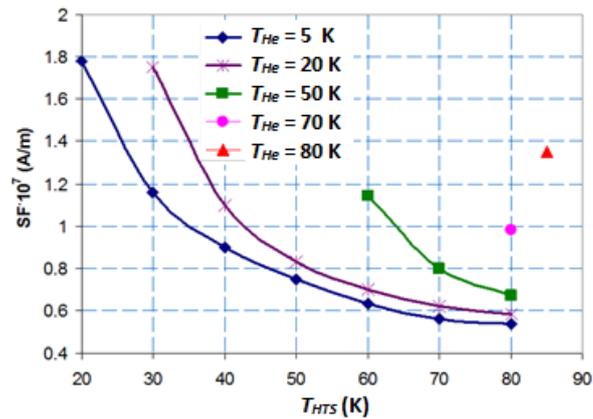

**Fig. 7:** Shape factor of the gas-cooled resistive part of a HTS lead as a function of the maximum operating temperature of the HTS ($T_{HTS}$) and of the inlet temperature of the helium gas ($T_{He}$). The geometry is calculated using the material properties of copper with RRR = 70.

The resistive parts of HTS leads are heat exchangers whose design can be based on the use of a large number of parallel wires or small tubes, or foil, soft-soldered to the copper terminals. However, for high current ratings, the alternative option based on a helical or meander flow of coolant presents several important advantages, namely: (i) reliable and consistent low resistance joints to the terminals can be easily made using standard techniques like electron-beam welding; (ii) the mass of the fins that creates the heat exchange area adds to the mass per unit length of the lead without increasing the heat conduction, thus increasing the thermal inertia of the lead, and time to thermal runaway, should the coolant flow be interrupted; and (iii) the heat exchanger is a well-defined component that can be readily specified and manufactured. For these reasons, recent large-scale applications have adopted the latter design for series production of current leads [11, 12].

The final optimization of the resistive part of a current lead designed for transferring high currents requires 2D and possibly 3D finite element calculations for the detailed definition of the geometry and the modelling of the thermal, electrical, and fluid dynamic performance [16].

## 3.2  HTS current leads: superconducting part

The thermal load ($Q_C$) of the superconducting part of a lead is given by the sum of the thermal conduction and of the power dissipation of the electrical joint at the cold end of the lead ($T_C$), where the HTS element is connected to a bus transferring the current to the superconducting device. Thermal conduction depends on the temperature gradient, on the conductor cross-section, on the length of the HTS element, and on the thermal conductivity of the material. $T_{\mathrm{HTS}}$ has to be chosen in accordance with Fig. 6. The final operating value, selected in the admitted temperature range, has to take into account the increased superconductor cross-section at higher temperatures, with consequent higher heat load and cost of the device. The length of the HTS part is for practical and economic reasons limited to a maximum of about 0.5 m [17, 12]. The HTS part can be conduction-cooled or self-cooled. Cooling with gas enables a significant reduction of the heat load into the liquid cryogen [9, 15].

HTS conductors for application in current leads must have low thermal conductivity. Protection requirements in case of resistive transition of the HTS impose the need for a shunt, parallel to the superconductor, able to transfer the current, after detection of a quench, during the time needed for discharging the electrical circuit. For this reason, the use of HTS leads may not offer any advantage in superconducting circuits that have very long time constants and that cannot provide a fast discharge in case of resistive transition of the HTS element (e.g. large detector magnets).

For protection reasons, HTS conductors with a metal matrix are preferred to bulk conductors, in particular for current leads transferring high currents. Bi-2223 has to date been the preferred choice for large-scale applications. It is produced as a multi-filamentary tape conductor with a silver alloy matrix. The typical cross-section of the tape is about 4 mm × 0.2 mm. The alloying of silver with gold enables a significant reduction of the thermal conductivity of the matrix with respect to pure silver (Fig. 8). Bi-2223 for use in current leads is produced specifically with a suitable gold alloyed matrix. The typical fraction of gold is 3% atomic.

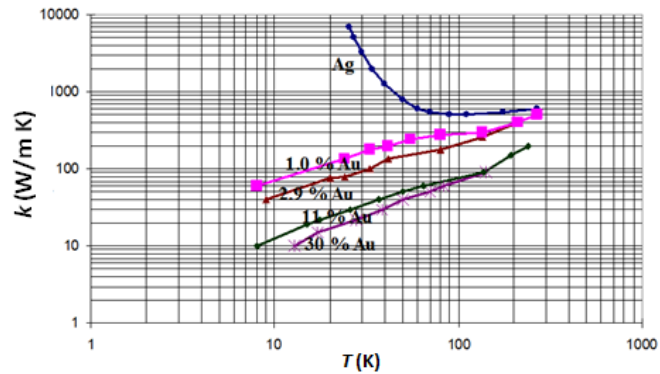

**Fig. 8:** Thermal conductivity of silver-gold matrix of Bi-2223 tape used in current leads [18]. Percentages of gold are atomic.

The large anisotropy of the Bi-2223 conductor and the consequent strong dependence of the critical current density on the field orientation shall be taken into account in the optimization of the geometry of the HTS part. Bi-2223 tapes are more sensitive to magnetic fields oriented perpendicular to the plane of the tape – a dependence that is particularly marked at higher temperatures. In the LHC current leads, for instance, up to nine Bi-2223 tapes are vacuum-soldered together to form rugged [19] superconducting components. The stacks are then disposed in a cylindrical configuration that orients the self-field parallel to the plane of the tapes [17].

YBCO tape is in principle also suitable for application to HTS current leads. It could be used without the copper stabilizer, which is usually soldered or electro-plated onto the tape, to limit the main thermal conduction to the contribution of the metallic substrate and buffer layers. An additional stabilizer with low thermal conductivity then has to be provided along the HTS part in order to meet the protection requirements that are defined by the characteristics of the electrical circuit.

## 4   HTS current leads: operation in pulsed mode

The previous sections describe the optimization of current leads operated at constant current. For the optimization of the resistive part of HTS leads operated in pulsed mode, the system of non-linear second-order partial differential equations expressing the heat balance between the lead and the gas in transient conditions in Eq. (12) must be solved numerically:

$$\frac{\partial}{\partial x}\left(k(T)A\frac{\partial T}{\partial x}\right) + \frac{\rho(T)I(t)^2}{A} - Ph(\vartheta)(T - \vartheta) = c_c(T)\,v_c A\frac{\partial T}{\partial t},$$

$$Ph(\vartheta)(T - \vartheta) - mc_p(\vartheta)\frac{\partial \vartheta}{\partial x} = c_p(\vartheta)\,v_g(\vartheta)A_g\frac{\partial \vartheta}{\partial t}. \qquad (12)$$

In Eq. (12), $t$ is the time, $c_c$, $v_c$ and $c_p$, $v_g$ are the specific heat and density of the conductor and coolant, respectively, and $A_g$ is the coolant cross-section. The system in Eq. (12) can be solved numerically taking as boundary conditions $T(x = 0,t)$ and $\vartheta(x = 0,t)$ and as initial conditions the temperature profile at $t = 0$ of the coolant and of the conductor ($T(x,t = 0)$ and $\vartheta(x,t = 0)$). For pulsed operation of superconducting systems having ramp rates of interest for some accelerators magnets – several kA s$^{-1}$ – the optimum performance is obtained by using a variable cross-section – larger in the top part of the lead [20].

AC losses in the Bi-2223 tape, which include hysteresis losses in the superconductor and eddy current losses in the silver alloy matrix, are a small fraction of the thermal conduction [21].

## 5   Current leads: operation requirements

A conventional self-cooled current lead and the gas-cooled resistive part of an HTS lead need to be protected against thermal runaway. The origin of the runaway can be a loss or a reduction of coolant flow rate due to operational issues. Also, conduction-cooled current leads need to be protected in case of improved performance and reduced cross-section via the use of intermediate heat sinks. Time constants in the case of thermal runaway are long – several seconds – and protection is given by the voltage measured across the devices that trigger a current discharge when a critical threshold is reached – typically of the order of 100 mV. The superconducting part of an HTS current lead has to be protected in case of resistive transition. The voltage measured across the HTS element serves as a quench protection signal.

The mass flow rate of self-cooled conventional leads or the gas-cooled resistive part of HTS leads needs to be controlled. In HTS leads, $T_{HTS}$ is the signal typically used for the control of the valve regulating the flow [22]. In conventional leads, a temperature sensor located toward the warm end is often used for control purposes.

## 6   Current leads for persistent mode operation

Persistent current operation is adopted in some applications, such as magnets for magnetic resonance imaging systems (MRI), where high stability of magnetic field (field decay $<10^{-8}$ $h^{-1}$) has to be guaranteed. After having been energized, the magnet is operated in a closed loop, with power supply turned off and current leads detached from the lower section. A superconducting switch inside the magnet, which is maintained above critical temperature during powering of the system, ensures the circuit continuity in the superconducting loop during persistent mode operation. To guarantee uniformity of current in time, resistive joints in the circuit must have sufficiently low resistance. Current leads for this type of application must be light, easily extractable, and equipped with retractors for automation and control of the break-away lead configuration. To meet these requirements, their optimization can be done for transient mode operation since transfer of current – and associated losses – are limited to the energizing phase of the system.

## 7   Buses and links

In an accelerator system, a current lead is connected at its cold end to a superconducting bus that provides the electrical link to the magnet. Superconducting buses also interconnect magnets that belong to the same electrical circuit. The bus is operating at a constant temperature in a liquid helium bath, and it consists of a Nb-Ti strand or cable designed for transporting the required current in the maximum field that the superconductor experiences. The design of a superconducting bus must take into account stability and protection requirements.

Resistive transitions in a bus can be induced by local increase in temperature due to transient disturbances, like movements or beam losses, or by resistive joints in the regions where splices between superconducting cables are made. If a normal zone appears and heat generation exceeds the available cooling, the normal zone propagates longitudinally along the bus with a speed given by the magnitude of the longitudinal quench velocity. Copper stabilizer is connected in parallel to the superconductor and acts as a bypass for the current during the discharge of the circuit that follows the detection of a resistive transition. The cross-section of the copper stabilizer, which has to be such as to limit the peak temperature reached during the transient, must be continuous along the bus length to avoid burnout of the cable when the resistive zone reaches the area of discontinuity. Such burnout can have major consequences if the bus is part of a circuit with high stored energy: an open circuit generates an electrical arc and a discharge of the energy in the cryostat cold mass. The LHC incident, in September 2008, was caused by a resistive splice in the main dipole Nb-Ti bus near an area with a discontinuity in the copper stabilizer [23].

A superconducting bus can be located in a dedicated cryostat and cooled by its own cryogenics. The system comprising the bus, the cryostat, and the cryogenic instrumentation and control equipment is called the superconducting link. Superconducting links made from HTS and $MgB_2$ superconductors and cooled by forced flow of helium gas are being developed [24, 25]. These links, which take advantage of properties of materials that remain superconducting at higher temperatures than classical superconductors, permit new concepts to be developed for power transmission systems. They have the potential of providing in the near future an attractive alternative way of delivering current to superconducting magnet circuits.